\documentclass[%
reprint,
superscriptaddress,
onecolumn,
 amsmath,amssymb,
 aps,
showkeys
]{revtex4-2}

\usepackage{graphicx}
\usepackage{epstopdf}
\usepackage{epsfig}
\usepackage{xcolor}
\usepackage{subcaption}
\usepackage{dcolumn}
\usepackage{bm}
\usepackage{booktabs}
\usepackage{mathrsfs}
\usepackage{bbold}

\usepackage{caption}
\makeatletter
\renewcommand\@makecaption[2]{%
  \par
  \vskip\abovecaptionskip
  \begingroup
   \small\rmfamily
    \begingroup
     \samepage
     \flushing
     \let\footnote\@footnotemark@gobble
     \@make@capt@title{#1}{#2}\par
    \endgroup
  \endgroup
  \vskip\belowcaptionskip
}
\makeatother

\begin{document}

\preprint{APS/123-QED}

\title{Going in circles: Slender body analysis of a self-propelling bent rod}

\author{Arkava Ganguly}
\author{Ankur Gupta}
 \email{Corresponding author:
 ankur.gupta@colorado.edu}
\affiliation{Department of Chemical and Biological Engineering, University of Colorado, Boulder}

\date{\today}

\begin{abstract}
We study the two-dimensional motion of a self-propelling asymmetric bent rod. By employing slender body theory and the Lorentz reciprocal theorem, we determine particle trajectories for different geometric configurations and arbitrary surface activities. Our analysis reveals that all particle trajectories can be mathematically expressed through the equation for a circle. The rotational speed of the particle dictates the frequency of the circular motion and the ratio of translational and rotational speeds describes the radius of the circular trajectory. We find that even for uniform surface activity, geometric asymmetry is sufficient to induce a self-propelling motion. Specifically, for uniform surface activity, we observe (i) when bent rod arm lengths are equal, the particle only translates, (ii) when the length of one arm is approximately four times the length of the other arm and the angle between the arms is approximately $\frac{\pi}{2}$, the rotational and translational speeds are at their maximum. We explain these trends by comparing the impact of geometry on the hydrodynamic resistance tensor and the active driving force. Overall, the results presented here quantify self-propulsion in composite-slender bodies and motivate future research into self-propulsion of highly asymmetric particles.
\end{abstract}

\maketitle

\section{Introduction} \label{sec:introduction}
Biological entities typically propel by the beating of cellular appendages like flagella or cilia in asymmetric, wave-like patterns \cite{elgeti2015physics, purcell1977life, berg1990chemotaxis, lauga2009hydrodynamics,lauga2006swimming,lauga2016bacterial}. To mimic biological motion, synthetic propellers have garnered attention due to their promising applications in medicine \cite{nelson2010microrobots,bunea2020recent,sundararajan2008catalytic,burdick2008synthetic}, microfluidic devices \cite{maggi2016self,sharan2021microfluidics}, environmental remediation \cite{sanchez2012spontaneous,guix2012superhydrophobic}, and the fabrication of self-repairing surfaces \cite{li2015self,pavel2021cooperative}.
Broadly speaking, there are two categories of propulsion mechanisms in synthetic particles. The first category of motion is externally actuated, where the propulsion is driven through an external field. For instance, magnetophoresis due to a magnetic field \cite{ghosh2009controlled,lim2011magnetophoresis,alnaimat2018microfluidics,alnaimat2018microfluidics,roure2020magnetization}, acoustic propulsion through ultrasound \cite{bertin2015propulsion,nadal2020acoustic,nadal2014asymmetric,xu2017ultrasound,mcneill2021purely,voss2022propulsion,ren2018two,voss2022orientation,mohanty2020contactless}, electrophoresis driven by constant electric fields \cite{khair2020migration,yee2018experimental,khair2022nonlinear,saad2018time,brooks2019shape,khair2018strong}, induced-charged electrophoresis due to AC electric fields \cite{gangwal2008induced,squires2004induced,squires2006breaking,bazant2010induced,khair2020breaking,brooks2018shape,lee2019directed,oren2020induced}, diffusiophoresis due to concentration gradients of solute(s) \cite{velegol2016origins,abecassis2008boosting,banerjee2016soluto,gupta2019diffusiophoretic,gupta2020diffusiophoresis,alessio2021diffusiophoresis}, and thermophoresis \cite{piazza2008thermophoresis,lin2017self,yang2011simulations} because of temperature gradients are all examples of externally driven phoretic motion. The second category of propulsion in synthetic particles is self-actuated, where the fields are generated by the particles themselves. Typical examples include self-diffusiophoresis and self-thermophoresis, among others \cite{gaspard2019stochastic,chen2018electrically,qin2017catalysis,jiang2010active,paxton2004catalytic,popescu2016self,wheat2010rapid,davis2022self,meredith2022chemical,meredith2020predator,kanso2019phoretic}. The focus of this work is self-diffusiophoresis, though the results outlined here are readily extended to self-thermophoresis as well. 
\par{} The most common example of self-diffusiophoresis reported in literature consists of a Janus sphere, where the motion is induced through an asymmetric reaction \cite{davis2022self,speck2019thermodynamic,chatterjee2018propulsion,popescu2018chemotaxis,zhou2018photochemically,nasouri2020exact}. However, several studies have argued that asymmetry in reaction is not a necessary requirement for self-diffusiophoresis. Instead, geometric asymmetries also induce a self-diffusiophoretic motion, even for a uniform surface activity.
Existing theoretical analyses have largely focused on specific particle geometries such as spheroidal \cite{poehnl2020axisymmetric,ebbens2011direct,shemi2018self,hsu2010diffusiophoresis,lisicki2018autophoretic} and cylindrical \cite{wang2013pnas,schnitzer2015osmotic}. However, the work by Shklyaev et al. \cite{shklyaev2014non} and Daddi-Moussa-Ider et al. \cite{daddi2021optimal} demonstrates that a perturbation to these shapes can modify the direction and speed of the propulsion. Clearly, geometry plays a key role in self-diffusiophoretic propulsion.

To go beyond these typical shapes, recent literature utilized slender body theory (SBT) \cite{hancock1953self,cox1970motion,batchelor1970slender} to predict the motion of self-diffusiophoretic particles. Schnitzer and Yariv \cite{schnitzer2015osmotic}, and Yariv \cite{yariv2019self} studied the motion of a straight slender rod with an arbitrary cross-section, arbitrary surface activity, and first-order reaction kinetics.  Poehnl and Uspal \cite{poehnl2021phoretic} investigated catalytic helical particles to obtain a good agreement between their SBT prediction and boundary element calculations.  Katsamba et al. \cite{katsamba2020slender,katsamba2022chemically} outlined a comprehensive SBT framework that can predict the motion for arbitrary surface activity and an arbitrary three-dimensional axisymmetric geometry.

While the studies described above advance our understanding of self-propulsion in slender bodies, they focus on a slender body with a single axis. In this work, we analyze the self-diffusiophoretic motion of a composite slender body, i.e., a bent-rod geometry. Our motivation to study a bent-rod is twofold. First, such an asymmetric geometry has been experimentally studied by K{\"u}mmel et al. \cite{kummel2013circular}, who reported a circular motion in L-shaped particles, which was later extended by Rao et al. \cite{rao2018self} who studied slender rods bent at different angles. Here, we describe the motion of similar geometries through SBT and do not invoke an external force and torque \cite{ten2015can}. Second, the hydrodynamics of a passive bent-rod have been studied in detail by Roggeveen and Stone \cite{roggeveen2022motion}. The authors calculate hydrodynamic mobility for such a geometry, which we utilize to predict the motion of a self-propelling bent rod. In section \ref{sec:framework}, we calculate the excess solute concentration and obtain the slip velocity at the particle surface. Next, we evaluate the particle motion by using the Lorentz reciprocal theorem \cite{masoud2019reciprocal,kim2013microhydrodynamics,poehnl2021phoretic}. Subsequently, we find that the particle trajectory is always circular. In section \ref{sec:results_discussion}, we validate our predictions with the experimental results of K\"{u}mmel et al. \cite{kummel2013circular} and obtain good quantitative agreement without any fitting parameters. Next, we investigate the scenario of uniform surface flux. Our model reveals the impact of geometry on the circular motion of particles. For specific geometric parameters, the translation-rotation coupling is counteracted by the rotation arising from surface activity, causing the particles to move in a straight line. We show that the translation and rotation speeds are maximum when one arm is approximately 4 times longer than the other and the arms are at right angles to each other. In section \ref{sec:conclusion}, we summarize our results, discuss the implications of our findings, and outline future directions.

\begin{figure}
    \centering
    \includegraphics[width=0.8\linewidth]{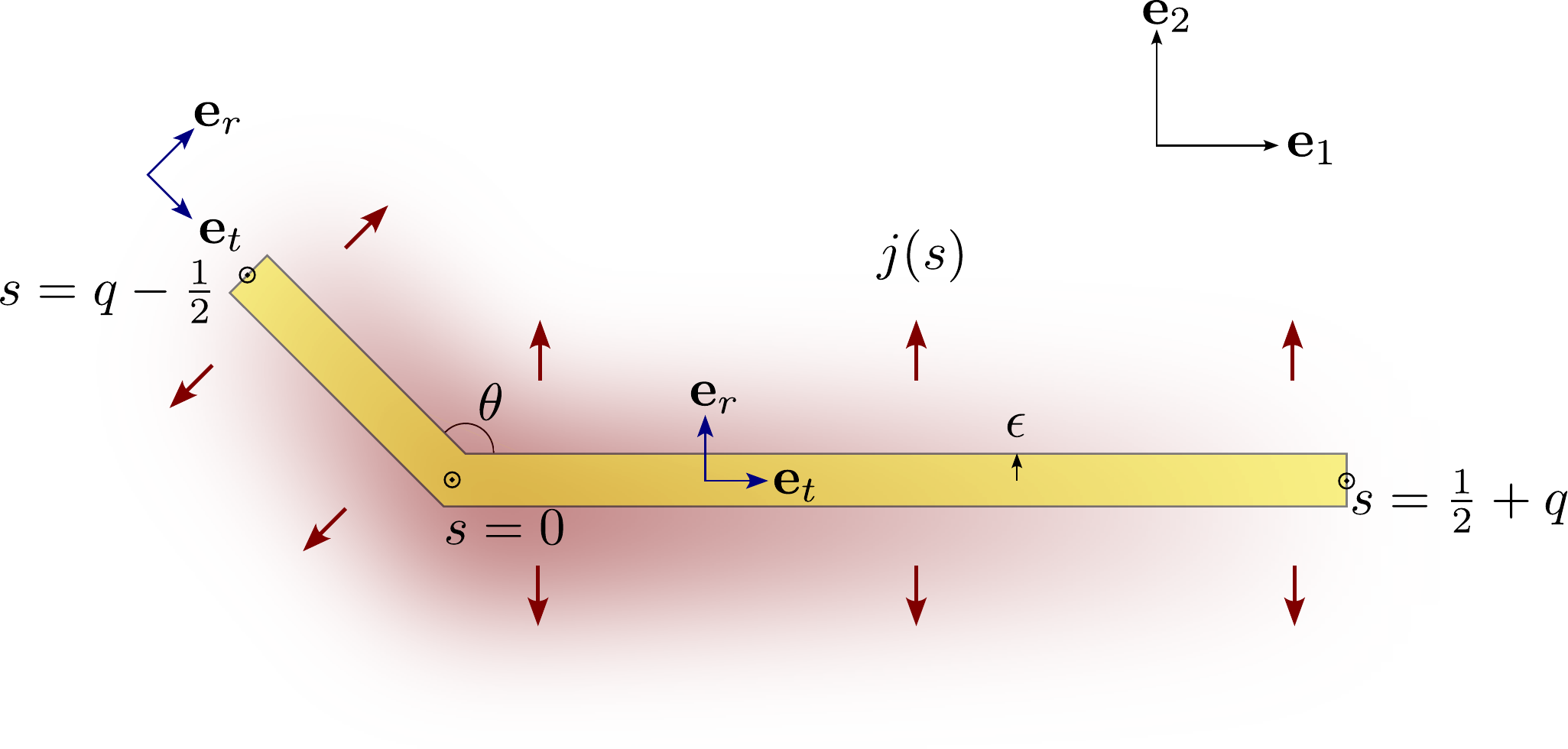}
    \caption{\textbf{Non-dimensional schematic of the problem setup.} We consider a rigid bent rod composed of two cylindrical arms of equal radius $a$,  aligned at an angle $\theta$. The length of the two arms are $ \left( \frac{1}{2} + q \right) \ell$ and $\left( \frac{1}{2} - q \right) \ell$. Therefore, the total length of the bent rod is $\ell$. We focus on the slender limit, i.e., $\epsilon = \frac{a}{\ell} \ll 1$. The rod self-propels due to solute flux on the rod in the $\mathbf{e}_1$-$\mathbf{e}_2$ plane and can rotate about the $\mathbf{e}_1 \times \mathbf{e}_2$ plane. $\mathbf{e}_{r}$-$\mathbf{e}_{t}$ represent the directions normal and tangential to the bent rod. To non-dimensionalize our problem setup, we scale all the lengths by $\ell$. $s$ represents the dimensionless coordinate along the rod.  $s=0$ is the hinge and $s = \frac{1}{2} \pm q$ are the ends of two arms. The dimensionless solute flux is represented by $j(s)$. Both $\mathbf{e}_1$-$\mathbf{e}_2$ and $\mathbf{e}_{r}$-$\mathbf{e}_{t}$ are define in reference frame of particle. $\mathbf{e}_1$ is defined such that is aligned with the positive arm, i.e., $ 0 \le s \le \frac{1}{2}+q$. $\mathbf{e}_2$ is perpendicular to $\mathbf{e}_1$. $\mathbf{e}_{r}$ and $\mathbf{e}_{t}$ are expressed as a function of $\mathbf{e}_1$ and $\mathbf{e}_2$; see Eq. (\ref{eq:tangential_vec}). The lab reference frame is given by $\mathbf{e}_x$-$\mathbf{e}_y$; see section \ref{sec: traj}.}
    \label{fig:1}
\end{figure}

\section{Theoretical framework} \label{sec:framework}
\subsection{Particle Geometry}\label{ssec:par_geom}
We follow the geometric description of a bent-rod outlined in Roggeveen and Stone \cite{roggeveen2022motion}. The bent-rod is composed of two cylindrical arms aligned at an angle $\theta$; see Fig. \ref{fig:1}. The lengths of the two arms are assumed to be $\left( \frac{1}{2} + q \right) \ell $ and $\left( \frac{1}{2} - q \right)\ell $, where $\ell$ is the total length and $q$ is the length asymmetry parameter. We note that $q \in \left[-\frac{1}{2},\frac{1}{2}\right]$. Both the arms are assumed to be of the same radius $a$ such that $\frac{a}{\ell} =\epsilon \ll 1$. The rod self-propels due to diffusiophoresis, induced by a surface reaction. We note that though the analysis presented here focuses on a diffusiophoretic process \cite{schnitzer2015osmotic,yariv2019self,morgan2014chemotaxis}, the results are also readily extendable to thermophoretic propulsion \cite{kummel2013circular,yang2011simulations}.

We non-dimensionalize the coordinate system by $\ell$. We introduce the arc-length parameter $s$ to describe the position along the centerline of the rod such that  $-\frac{1}{2} + q \le s \le \frac{1}{2} + q $. $s=0$ represents the hinge, whereas $s=q \pm \frac{1}{2}$ denote the end of the two arms.  For consistency, we refer to the arm where $ 0 \le s \le q + \frac{1}{2}$ as the positive arm and the arm where  $ -\frac{1}{2} + q \le s < 0 $ as the negative arm. The shape of the bent rod is thus dictated by $q$ and $\theta$.

We assume that the rod only propels in the $\mathbf{e}_1$-$\mathbf{e}_2$ plane and can rotate about the $\mathbf{e}_3 = \mathbf{e}_1 \times \mathbf{e}_2$ axis. The direction $\mathbf{e}_1$ is always assumed to be aligned with the positive arm. For convenience, we also define $\mathbf{e}_t$ and $\mathbf{e}_r$ as the tangential and normal directions to the rod, respectively, such that

\begin{subequations}
\begin{equation}
    \mathbf{e}_t=\begin{cases}
    - \cos \theta \ \mathbf{e}_1 - \sin \theta \ \mathbf{e}_2 & s < 0 \\
    \mathbf{e}_1 & s \geq 0
    \end{cases},\label{eq:tangential_vec}
\end{equation}

\begin{equation}
    \mathbf{e}_r = \begin{cases}
    -\sin \theta \ \mathbf{e}_1 + \cos \theta \ \mathbf{e}_2 & s < 0  \\
    \mathbf{e}_2 & s \geq 0
    \end{cases}.\label{eq:radial_vec}
\end{equation}
\label{eq:basis_vec}
\end{subequations}

\noindent Note that both $\mathbf{e}_1$-$\mathbf{e}_2$ and $\mathbf{e}_t$-$\mathbf{e_r}$ are in the particle frame of reference and moves with the particle. In section \ref{sec: traj}, we define $\mathbf{e}_x$-$\mathbf{e}_y$ as our universal frame of reference to obtain equations for the particle trajectories; see Eq. (\ref{eq:univ_vel}). The position of a point on the particle centerline is $\mathbf{x}_h(s) = s \mathbf{e}_t$. The center of mass of the bent-rod is denoted as $\mathbf{x}_{\rm com}$. To model self-propulsion through catalytic activity, we follow the common practice in literature \cite{schnitzer2015osmotic,yariv2019self,poehnl2021phoretic,popescu2016self,anderson1989colloid, golestanian2007designing,golestanian2005propulsion}, and assume a solute flux $j(s)$ on the particle surface (the mathematical definition of $j(s)$ is provided later). The induced translation and rotation velocities of the particle are denoted by $\mathbf{U}$ and $\mathbf{\Omega}$, respectively. The objective of this paper is to determine the particle trajectory in the limit $\epsilon \ll 1$ for a given $q$, $\theta$, and $j(s)$.
The limit $\epsilon \ll 1$ enables us to invoke first-order slender body theory to evaluate $\mathbf{U}$ and $\mathbf{\Omega}$. We follow the approach outlined in Schnitzer and Yariv \cite{schnitzer2015osmotic} and Poehnl and Uspal \cite{poehnl2021phoretic} to obtain the excess solute concentration profile and effective slip velocity.  Next, we employ the geometric resistance coefficients obtained from Roggeven and Stone \cite{roggeveen2022motion} and use the Lorentz reciprocal theorem to obtain the particle trajectory for a self-propelling bent rod.
Since we utilize first-order slender body theory, we superpose the concentration and hydrodynamic effects of the two arms and neglect the higher-order interactions between them. Therefore, our analysis becomes less applicable for cases where the interaction between the arms become important. We also acknowledge that our analysis ignores the circumferential variations in the solute flux, discussed in-depth by Kastamba et al. \cite{katsamba2020slender,katsamba2022chemically}.

\subsection{Concentration Profile} \label{ssec:conc_prof}
We seek to evaluate the concentration of the solute at the particle surface for a given geometry and surface flux. To do so, we define dimensionless surface flux $j(s)$ scaled by reference flux $J_{\textrm{ref}}$. We define $c(s,r)$ to be dimensionless solute concentration, scaled by $\frac{a J_{\textrm{ref}}}{D}$, where $D$ is the solute diffusivity.  Mathematically, our objective is to evaluate  concentration at the slip plane $c_s(s)$ for a given $q$, $\theta$ and $j(s)$. We note that $c_s(s)$ is equivalent to the surface concentration from the outer solution $c^{\textrm{out}}(s,\epsilon)$ \cite{poehnl2021phoretic,schnitzer2015osmotic,bender1999advanced} (also see Appendix A). We define the P{\'e}clet number of the rod as Pe$=\frac{U_{\textrm{ref}} \ell}{D}$, where $U_{\textrm{ref}}$ is a typical velocity scale. For representative values, we focus on ref. \cite{dey2013ph}. Here, catalytic spheres with $2 \mu$m diameters were driven in H$_2$O$_2$ solutions.  $D=O(10^{-8})$ m$^2$/s, $\ell_{\textrm{ref}}=O(10^{-6})$ m, and $U_{\textrm{ref}}=O(10^{-5})$ m/s. Therefore,  Pe = $O(10^{-3})$. This helps us justify neglecting convection and unsteady terms in Eq. \eqref{Eq: cons_eqn} (additional justification is provided below Eq. (\ref{eq:steady-state trajectory})). Therefore, we write

\begin{equation}
     \nabla^2 c = 0, r \ge \epsilon.
     \label{Eq: cons_eqn}
 \end{equation}
 
\noindent The diffusiophoretic activity is represented with a surface flux boundary condition,
\begin{subequations}
\begin{equation}
    - \epsilon \hat{\mathbf{n}} \cdot \nabla c  = j(s),\ r=\epsilon,
    \label{Eq: flux_cond}
\end{equation}
where $\hat{\mathbf{n}}$ is the surface normal vector. The far field boundary condition for solute concentration reads
\begin{equation}
    c =0, r \rightarrow \infty.
        \label{Eq: farfield}
\end{equation}
\end{subequations}
Since $\epsilon \ll 1$, we use boundary-layer theory \cite{bender1999advanced} to evaluate $c_s(s)$. As outlined in Appendix A, we divide the fluid volume into an inner and an outer region in the radial direction $\mathbf{e}_r$. In the inner region, we stretch the coordinates such that $\rho = r \epsilon^{-1}$ and evaluate $c^{\textrm{in}}(s, \rho)$ with the boundary condition in Eq. \eqref{Eq: flux_cond}. Subsequently, in the outer region, we evaluate $c^{\textrm{out}}(s, r)$ as a line integral of diffusive sources of strength $\alpha(s)$. We determine $\alpha(s)$ via an asymptotic matching $c_{\textrm{in}}(s, \infty) = c_{\textrm{out}}(s, \epsilon)$. From the leading order behavior, we obtain $\alpha(s) = \frac{j(s)}{2}$, which gives

\begin{equation}
        c_s(s)=\frac{1}{2}\displaystyle \int_{q-\frac{1}{2}}^{q+\frac{1}{2}} \frac{j(s')}{\left|\mathbf{x}_h(s) + \epsilon \mathbf{e}_r(s) -\mathbf{x}_h(s')\right|} \rm d s'. \label{eq:unifrm_con_sol}
\end{equation}
\noindent The expression for $c_s(s)$ in Eq. \eqref{eq:unifrm_con_sol} is consistent with the results of Schnitzer and Yariv \cite{schnitzer2015osmotic} and Poehnl and Uspal \cite{poehnl2021phoretic}. Our analysis deviates from previous studies since we account for a composite slender body where $\mathbf{e}_t$ and $\mathbf{e}_r$ are different for the two arms. We define the concentration differences between the junction and the respective end-points as $\Delta c^{q+\frac{1}{2}}_0 = c_s \left(q+\frac{1}{2}\right) - c_s (0)$, and $\Delta c_{q-\frac{1}{2}}^0 = c_s (0) - c_s \left(q-\frac{1}{2}\right)$. As we show later,  $\Delta c^{q+\frac{1}{2}}_0$ and $\Delta c_{q-\frac{1}{2}}^0$ drive particle motion. \par{}

\subsection{Particle Velocity} \label{ssec:par_vel}
We define a dimensionless fluid velocity $\mathbf{u}$ around the particle, scaled by a reference velocity $U_{\rm ref}$, where $U_{\textrm{ref}} = \frac{k_B T a J_{\textrm{ref}} \lambda^2}{ \mu D \ell}$, where $k_B$ is the Boltzmann constant, $T$ is the absolute temperature, and $\lambda$ is the interaction length scale. Typically, $\lambda_{\textrm{ref}} = O(1)-O(10)$ nm \cite{anderson1989colloid}. Following the analyses in literature \cite{schnitzer2015osmotic,golestanian2007designing,katsamba2020slender}, we represent the interaction between the solute and the particle through a diffusiophoretic slip velocity \cite{anderson1982motion,anderson1989colloid}, such that
\begin{equation}
    \mathbf{u}_{\textrm{slip}}=M(s)\left(\mathbf{I}-\mathbf{e}_r\mathbf{e}_r\right)\cdot\nabla c_s = M(s) \frac{d c_s}{d s} \mathbf{e}_t, \label{eq:slip_velocity}
\end{equation}
where $M(s)$ is a non-dimensional lumped mobility parameter scaled by $\frac{k_B T \lambda^2}{\mu}$. For simplicity, we consider $M(s) = 1$. To justify $M(s)=1$, we note $M=O \left(\frac{ U_{\textrm{ref}} }{ k_B T c_{\textrm{ref}} \lambda^2/(\mu \ell_{\textrm{ref}})} \right)$. For $k_B = O(10^{-23}) \ $J/K, $T=300$ K, $c_{\textrm{ref}}=O(10^{24})$ m$^{-3}$, $\lambda=O(10^{-9})$ m \cite{anderson1989colloid}, $\mu=O(10^{-3})$ Pa$\cdot$s, $\ell_{\textrm{ref}}=O(10^{-6})$ m, we get $M=O(1)$. Before proceeding with hydrodynamic calculations, we highlight that while Eq. (\ref{eq:unifrm_con_sol}) is derived for a diffusiophoretic system, the results can also be extended to a thermophoretic system. Specifically, the surface concentration $c_s$ can be replaced by the dimensionless surface temperature (appropriately scaled by subtracting far-field temperature), and the point sources of solute flux can be replaced by point sources of heat flux. \par{} The fluid velocity at the particle surface for a phoretic particle with translational velocity $\mathbf{U}$ (scaled by $U_{\rm ref}$) and rotational velocity $\mathbf{\Omega}$ (scaled by $\frac{U_{\rm ref}}{\ell}$) is described as

\begin{equation}
    \mathbf{u}(s,\epsilon) = \mathbf{U}+\mathbf{\Omega}\times \left(\mathbf{x}_h - \mathbf{x}_{\textrm{com}}\right) + \mathbf{u}_{\textrm{slip}}. \label{eq:active_vel_bnd_cnd}
\end{equation}

\noindent In addition, the fluid velocity vanishes in the far-field, $\mathbf{u}\left(s,r \to \infty\right) \to 0$. Instead of solving for the velocity field, we employ the Lorentz reciprocal theorem to estimate $\mathbf{U}$ and $\mathbf{\Omega}$ \cite{masoud2019reciprocal,popescu2016self}. To this end, we relate the fluid velocity $\mathbf{u}$ to an auxiliary Stokes flow around the same particle geometry, with velocity $\mathbf{u}^j$ as follows \cite{poehnl2021phoretic},

\begin{equation}
    \displaystyle \int_S \mathbf{u}\cdot\mathbf{\sigma}^j\cdot\hat{\mathbf{n}} \  \rm d S = \displaystyle \int_S \mathbf{u}^j\cdot\mathbf{\sigma}\cdot\hat{\mathbf{n}} \ \rm d S, \label{eq:reciprocal_relation}
\end{equation}
 \noindent where the surface stresses $\mathbf{\sigma}$ and $\mathbf{\sigma}^j$ are scaled by $\frac{4 \pi \mu U_{\rm ref}}{\ell}$. The auxiliary problem has a no-slip condition at the rigid particle boundary, or 
 \begin{equation}
     \mathbf{u}^j(s,\epsilon) = \mathbf{U}^j+\mathbf{\Omega}^j \times \left(\mathbf{x}_h - \mathbf{x}_{\textrm{com}}\right). \label{eq:pas_vel_bnd_cnd}
 \end{equation}
 
 \noindent The fluid velocity vanishes in the far-field, $\mathbf{u}^j (s, r \rightarrow \infty) = \mathbf{u}^{j, \infty} \to \mathbf{0}$. Since the particle is constrained to move in the $\mathbf{e}_1$-$\mathbf{e}_2$ plane, $\mathbf{U}=U_1 \mathbf{e}_1 + U_2 \mathbf{e}_2$ and $\mathbf{\Omega}= \Omega_3 \mathbf{e}_3$, implying we need three different auxiliary problems to obtain $\mathbf{U}$ and $\mathbf{\Omega}$. We consider the auxiliary problems to be classified as $j \in \{1,2,3\}$ such that

\begin{subequations}
\label{eq:auxilliary}
\begin{eqnarray}
    & \mathbf{U}^1 = V_0 \mathbf{e}_1,\ \mathbf{\Omega}^{1} = \mathbf{0}, \\[10pt]
    & \mathbf{U}^2 = V_0 \mathbf{e}_2,\ \mathbf{\Omega^{2}} = \mathbf{0}, \\[10pt]
    & \mathbf{U}^3 = \mathbf{0},\ \mathbf{\Omega^{3}} = \Omega_0 \mathbf{e}_3.
\end{eqnarray}
\end{subequations}

 We substitute Eqs. \eqref{eq:active_vel_bnd_cnd} and \eqref{eq:pas_vel_bnd_cnd} into Eq. \eqref{eq:reciprocal_relation} to obtain
 
 \begin{equation}
    \begin{aligned}
    \mathbf{U}\cdot&\displaystyle\int_S \mathbf{\sigma^j}\cdot\hat{\mathbf{n}} \ \rm dS \it + \mathbf{\Omega}\cdot\int_S \left(\mathbf{x}_h(s) - \mathbf{x}_{\rm com}\right) \times \mathbf{\sigma}^j\cdot\hat{\mathbf{n}} \  \rm dS + \it \int_S \mathbf{u}_{\rm slip}\it \cdot\sigma^j\cdot\hat{\mathbf{n}} \ \rm dS \\
    &\it = \mathbf{U}^j\cdot\int_S \mathbf{\sigma}\cdot\hat{\mathbf{n}} \ \rm dS + \it \Omega^j\cdot\int_S \left(\mathbf{x}_h(s)-\mathbf{x}_{\rm com}\right) \it \times \mathbf{\sigma}\cdot\hat{\mathbf{n}} \ \rm dS.
    \end{aligned}\label{eq:active_passive_solution}
\end{equation}

\noindent Since a self-propelling particle is force and torque free \cite{popescu2016self,masoud2019reciprocal,poehnl2021phoretic},  the right hand side of Eq. \eqref{eq:active_passive_solution} vanishes. Further, we can represent the surface integral of the hydrodynamic stress $\mathbf{\sigma}^j$ in terms of the line integral of the hydrodynamic force density $\mathbf{f}^j$, $\mathbf{\sigma}^j \cdot \hat{\mathbf{n}} \ \rm{dS} = \it \mathbf{f}^j \ \rm ds$. In slender body theory \cite{kim2013microhydrodynamics,batchelor1970slender}, the force density $\mathbf{f}^j$ (scaled by $4 \pi \mu U_{\rm ref}$) is defined as,
\begin{equation}
    \mathbf{f}^j(s)=-\mathscr{A}\left(\mathbf{I}-\frac{1}{2}\mathbf{e}_t \mathbf{e}_t\right)\cdot \left(\mathbf{u}^j-\mathbf{u}^{j,\infty}\right),\label{eq:hydrodynamic_force_density}
\end{equation}

\noindent where $\mathscr{A} = \frac{1}{\ln \left(1/\epsilon\right)}$. The first two integrals in Eq. \eqref{eq:active_passive_solution} are the non-dimensional hydrodynamic force $\mathbf{F}^j$ and torque $\mathbf{T}^j$ on the particle in the auxiliary flow problem. We thus obtain

\begin{equation}
    \mathbf{U}\cdot\mathbf{F}^j+\mathbf{\Omega}\cdot\mathbf{T}^j + \displaystyle \int_{q-\frac{1}{2}}^{q+\frac{1}{2}} \mathbf{u}_{\textrm{slip}} \cdot \mathbf{f}^j(s) \ \rm d s = 0. \label{eq:linear_equation}
\end{equation}

\noindent To evaluate $\mathbf{F}^j$ and $\mathbf{T}^j$ for given $\mathbf{U}^j$ and $\mathbf{\Omega}^j$, we invoke the results of Roggeveen and Stone \cite{roggeveen2022motion}. The resistance coefficients, as provided in Roggeveen and Stone \cite{roggeveen2022motion}, $\mathbf{A}$, $\mathbf{B}$, and $\mathbf{C}$ (see Appendix B), allow us to write 

\begin{equation}
    \begin{bmatrix}
      F^j_1 \\
      F^j_2 \\
      T^j_3
    \end{bmatrix} = \begin{bmatrix}
      A_{11} & A_{12} & \Tilde{B}_{13} \\
      A_{12} & A_{22} & \Tilde{B}_{23} \\
      \Tilde{B}_{13} & \Tilde{B}_{23} & C_{33}
    \end{bmatrix}\begin{bmatrix}
      U^j_1 \\
      U^j_2 \\
      \Omega^j_3
    \end{bmatrix}.\label{eq:hydrodynamic_resistance}
\end{equation}

\noindent Substituting Eqs. (\ref{eq:auxilliary}), (\ref{eq:hydrodynamic_force_density}), and (\ref{eq:hydrodynamic_resistance}) in Eq. (\ref{eq:linear_equation}),  we obtain a system of linear equations to calculate phoretic motion of the particle

\begin{equation}
    \begin{aligned}
    \begin{bmatrix}
    U_1 \\ U_2 \\ \Omega_3
    \end{bmatrix} = &\mathscr{A}\mathbf{R}^{-1}\begin{bmatrix}
        \displaystyle \int_{\left(q-\frac{1}{2}\right)}^{\left(q+\frac{1}{2}\right)} \mathbf{u}_{\rm slip}\cdot(\mathbf{I}-\frac{1}{2}\mathbf{e}_t\mathbf{e}_t)\cdot\mathbf{e}_1 \  \rm d s \\
        \displaystyle \int_{\left(q-\frac{1}{2}\right)}^{\left(q+\frac{1}{2}\right)} \mathbf{u}_{\rm slip}\cdot(\mathbf{I}-\frac{1}{2}\mathbf{e}_t\mathbf{e}_t)\cdot\mathbf{e}_2 \ \rm d s \\
        \displaystyle \int_{\left(q-\frac{1}{2}\right)}^{\left(q+\frac{1}{2}\right)} \mathbf{u}_{\rm slip}\cdot(\mathbf{I}-\frac{1}{2}\mathbf{e}_t\mathbf{e}_t)\cdot\mathbf{e}_3 \times\left(\mathbf{x}_h\left(s\right)-\mathbf{x}_{\rm com}\right) \  \rm d s
    \end{bmatrix},
    \end{aligned}\label{eq:linear_sys_hydrodynamic}
\end{equation}

\noindent where $\mathbf{R}$ is the scaled resistance matrix. Note that the velocities are zero if $\mathbf{u}_{\rm slip}=0$. Substituting the expression of $\mathbf{u}_{\rm slip}$ from Eq. \eqref{eq:slip_velocity} in the above equation and integrating, we obtain

\begin{equation}
\begin{bmatrix}
    U_1 \\ U_2 \\ \Omega_3
  \end{bmatrix}=\mathscr{A} \begin{bmatrix}
   a_{11} & a_{12} & \Tilde{b}_{13} \\
   a_{12} & a_{22} & \Tilde{b}_{23} \\
   \Tilde{b}_{13} & \Tilde{b}_{23} & c_{33}
  \end{bmatrix}
  \begin{bmatrix}
    \frac{1}{2} \Delta c_0^{q+\frac{1}{2}} - \frac{\cos \theta}{2} \Delta c_{q-\frac{1}{2}}^0 \\
    - \frac{\sin \theta}{2} \Delta c_{q-\frac{1}{2}}^0 \\
    \frac{1}{4}\left(q-\frac{1}{2}\right)^2 \sin \theta \Delta c_0^{q+\frac{1}{2}} + \frac{1}{4}\left(q+\frac{1}{2}\right)^2 \sin \theta \Delta c_{q-\frac{1}{2}}^0
    \end{bmatrix}. \label{eq:mobility_relation}
\end{equation}

\noindent The coefficients of the mobility, i.e., $\mathbf{a}$, $\mathbf{b}$, and $\mathbf{c}$, can be obtained by inverting the resistance matrix in Eq. \eqref{eq:hydrodynamic_resistance} (see Appendix B).  We underscore that Eq. \eqref{eq:mobility_relation} is valid for an arbitrary $j(s)$, which modifies the values of $\Delta c_0^{q+\frac{1}{2}}$ and $\Delta c_{q-\frac{1}{2}}^0$. Additionally,  Eq. \eqref{eq:mobility_relation} demonstrates that $\mathbf{U}$ and $\mathbf{\Omega}$ are dictated by concentration differences between the end-points and the hinge, i.e., $\Delta c_0^{q+\frac{1}{2}}$ and $\Delta c_{q-\frac{1}{2}}^0$, and the spatial variation of the concentration profile does not influence the particle motion.

\subsection{Particle Trajectories}
\label{sec: traj}
$\mathbf{U}$ and $\mathbf{\Omega}$ given by Eq. \eqref{eq:mobility_relation} are in the reference frame of the particle, i.e., $\mathbf{e}_1$-$\mathbf{e}_2$. To obtain an equation of motion for the center of mass of the particle, i.e., $\left[x(t), y(t)\right]$, in a universal frame of reference $\mathbf{e}_x$-$\mathbf{e}_y$, we employ the rotation matrix as 

\begin{equation}
\frac{d}{dt}
    \begin{bmatrix}
        \vspace{0.08in} x(t) \\ \vspace{0.08in} y(t) \\ \phi(t)
    \end{bmatrix} = \begin{bmatrix}
        \vspace{0.08in} \cos \phi(t) & -\sin \phi(t) & 0 \\
        \vspace{0.08in} \sin \phi(t) & \cos \phi(t) & 0 \\
        0 & 0 & 1
    \end{bmatrix} \begin{bmatrix}
        \vspace{0.08in} U_1 \\ \vspace{0.08in} U_2 \\ \Omega_3
    \end{bmatrix}, \label{eq:univ_vel}
\end{equation}
where $\phi$ is the angle between $\mathbf{e}_1$ and $\mathbf{e}_x$, and $t$ is dimensionless time scaled by $\ell/U_{\textrm{ref}}$. Without any loss of generality, we define $x(0)=y(0)=\phi(0)=0$. By  integrating  (\ref{eq:univ_vel})  in time, we get

\begin{equation}
    \begin{aligned}
      \left(x(t) - \mathscr{X}\right)^2 + \left(y(t)-\mathscr{Y}\right)^2 = \mathscr{R}^2
    \end{aligned},\label{eq:steady-state trajectory}
\end{equation}
\noindent where $\mathscr{X}=- \frac{U_2}{\Omega_3} $ and $\mathscr{Y}=\frac{U_1}{\Omega_3}$. Eq. (\ref{eq:steady-state trajectory}) demonstrates that the particle moves in a circular trajectory where the center of the trajectory depends on the particle velocities. The radius of curvature of the trajectory is given by  $\mathscr{R}=\sqrt{\frac{U_1^2+U_2^2}{\Omega_3^2}}$. The turn frequency of the particle is simply $f=\frac{\Omega_3}{2\pi}$. We note that Eq. (\ref{eq:steady-state trajectory}) implicitly assumes that diffusion is significantly faster than the particle motion for Pe = $O(10^{-3})$, as previously discussed in section  \ref{ssec:conc_prof}. At this point, we will like to further justify the pseudosteady state assumption. Since our predictions suggest a circular trajectory, one might argue that the continuous release of solute could leave a trace behind that can build up over time. To quantify this effect, we briefly restore dimensions and perform an order of magnitude analysis. The total amount of solute released by the particle for time $\tau$ is given by $m \sim J_{0} (2 \pi a \ell) \tau$, where $J_0$ is the dimensional surface flux. The trace concentration $C_{\textrm{trace}}$ can be estimated as $C_{\textrm{trace}} \sim m/\left(\frac{\pi}{6} \delta^3 \right)$, where $\delta \sim \sqrt{4 D \tau}$ is the diffusive boundary layer thickness. Upon rearrangement, it is easy to obtain that $\frac{C_{\textrm{trace}}}{C_{\textrm{ref}}}\sim \frac{3}{2} j_0 \sqrt{ \frac{\ell^2}{D \tau} }$. Since Eq. (\ref{eq:steady-state trajectory}) is calculated for $\tau \sim \ell/U_{\textrm{ref}}$,  $\frac{C_{\textrm{trace}}}{C_{\textrm{ref}}} \sim \frac{3}{2} j_0 \textrm{Pe}^{1/2}  \ll c_s \sim O \left(j_0 \log \left( \frac{1}{\epsilon} \right) \right)$. Therefore, solute concentration generated due to surface activity is significantly higher than the trace solute concentration, which enables us to employ the quasi-steady state assumption.

\section{Results and Discussion} \label{sec:results_discussion}

\begin{figure}[t]
    \centering
    \includegraphics[width=0.85 \textwidth]{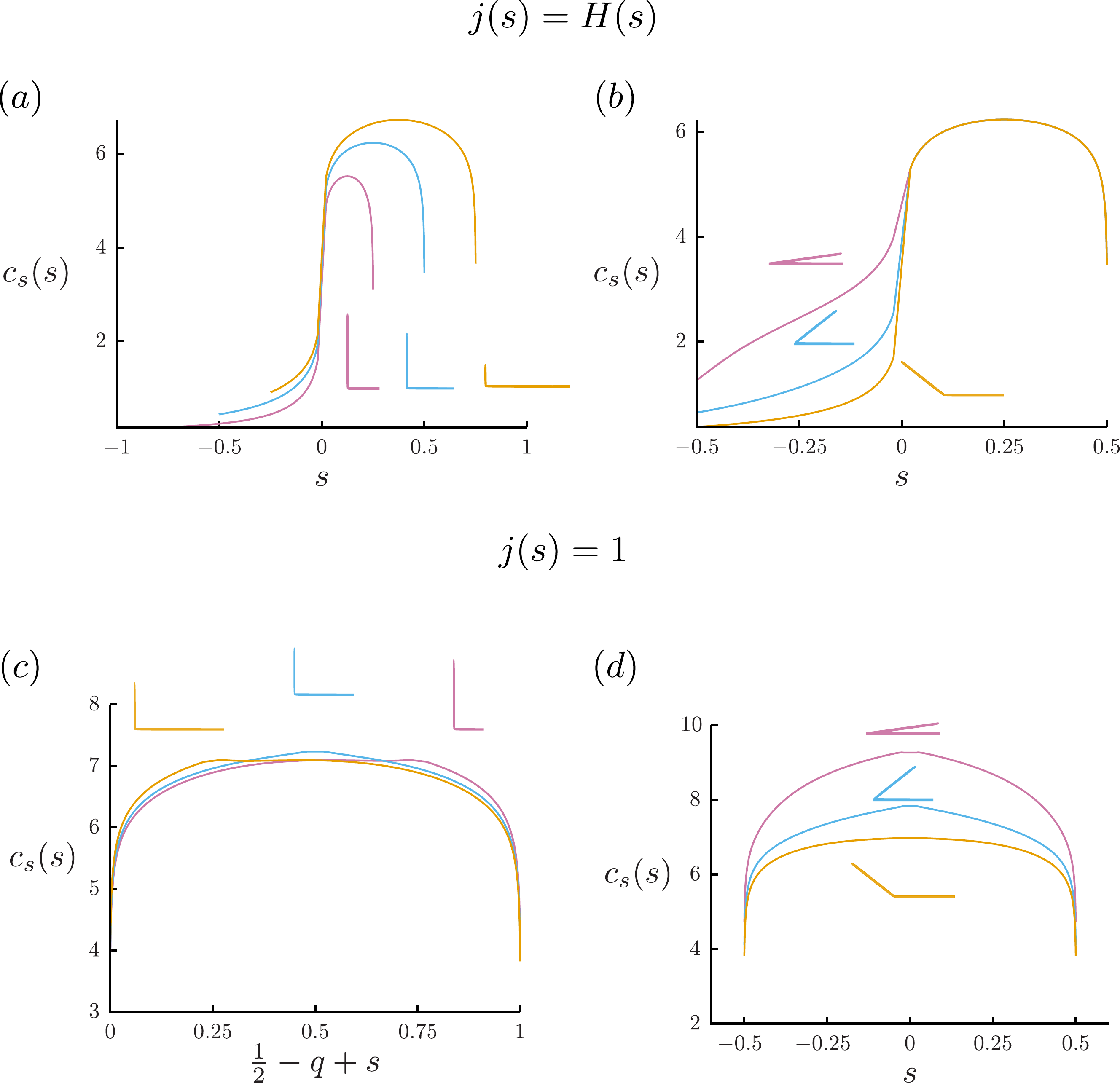}
    \caption{ \textbf{Surface concentration $c_s(s)$ for different combinations of $q$, $\theta$ and $j(s)$.}  (a, b) $j(s)=H(s)$ and (c,d) $j(s)=1$. (a, c) $\theta=\frac{\pi}{2}$ and $q=-0.25$ (pink), $q=0$ (blue), and $q=0.25$ (yellow). (b, d)  $q=0$ and $\theta=\frac{\pi}{18}$ (pink), $\theta = \frac{\pi}{4}$ (blue), and $\theta = \frac{3\pi}{4}$ (yellow).  $\epsilon=10^{-3}$ for all cases.}
    \label{fig:2}
\end{figure}

\subsection{Validation} For a given $j(s)$, $q$, and $\theta$, Eqs. (\ref{eq:unifrm_con_sol}), (\ref{eq:mobility_relation}), and (\ref{eq:steady-state trajectory}) can be employed to calculate the particle trajectory. To test the validity of our approach, we compare our predictions with the experimental trajectories of a L-shaped thermophoretic particle, as described in K\"{u}mmel et al. \cite{kummel2013circular}. We note that the L-shaped particle utilized by the authors does not consist of cylindrical arms. However, at the slender limit, the cross-sectional geometry will have a minor impact on the calculations. Therefore, we are able to use the methodology described above. 
\par{}   The geometry of L-shaped particle in  K\"{u}mmel et al. \cite{kummel2013circular} yields $\ell=12$ \ $\mu$m, $\epsilon=0.125$, $q=-\frac{1}{8}$, and $\theta=\frac{\pi}{2}$. Since only the positive arm of the L-shaped geometry in K\"{u}mmel et al. \cite{kummel2013circular} is thermophoretically active, $j(s)=H(s)$, where $H$ is the Heaviside function. Substituting geometric values and $j(s)=H(s)$ in Eq. (\ref{eq:unifrm_con_sol}), we evaluate $\mathbf{U}$ and $\mathbf{\Omega}$ by using Eq. (\ref{eq:mobility_relation}), and obtain the trajectory through Eq. (\ref{eq:steady-state trajectory}).  First, we are able to recover the circular trajectory observed experimentally by  K\"{u}mmel et al. \cite{kummel2013circular}. Further, we obtain $\ell \mathscr{R}=8.1 \ \mu$m, which is in surprisingly good quantitative agreement with the experimentally reported radius of $7.5 \ \mu$m. We emphasize that no fitting parameter was employed for this comparison. Finally, while the values of $\mathbf{U}$ and $\mathbf{\Omega}$ are dependent on the magnitude of $j$ and $\epsilon$, the value of $\mathscr{R}$ is independent of the magnitude $j$ and $\epsilon$, and is only a function of $q$ and $\theta$. Physically, this result implies that the radius of the circular trajectory of the particle is less sensitive to the magnitude of the reactive flux and the slenderness of the geometry. However, both the speed of the particle and the frequency of rotation increase with an increase in slenderness, i.e., lower $\epsilon$, and an increase in reactive flux. 

\par{} Interestingly, for $j(s)=H(s)$, $\Delta c_{0}^{q+\frac{1}{2}}=0$  and $\Delta c_{q-\frac{1}{2}}^{0} \neq 0$, see Fig. \ref{fig:2}(a,b). This finding implies that for $j(s)= H(s)$, the concentration about the active arm is symmetric and the concentration difference across it is negligible. The $3 \times 1$ matrix in the RHS of Eq. (\ref{eq:mobility_relation}) is the effective force and torque generated by the active propulsion mechanism. Since $\Delta c_0^{q+\frac{1}{2}}=0$ for $j(s)=H(s)$, the effective force and torque are driven solely by the diffusive tail on the negative arm, i.e., the passive arm. Furthermore, our analysis reveals that the effective force is parallel to the passive arm.  This result is in contrast to the findings of K\"{u}mmel et al. \cite{kummel2013circular}, who assumed that a force is applied perpendicular to the active arm.  In fact, our analysis enables us to move away from the artificial force and torque approach \cite{ten2015can}, as we are able to evaluate the effective force and torque by utilizing the Lorentz reciprocal theorem. Our analysis suggests that, for multiple arms, the effective force $\mathbf{F}_{\textrm{eff}} \propto \sum_i \Delta c_0^{i} \mathbf{e}_{ti}$, where $\Delta c_0^{i}$ is the concentration difference along the $i^{\textrm{th}}$ arm and $\mathbf{e}_{ti}$ is the parallel vector to the arm. 
\begin{figure}[t]
    \centering
    \includegraphics[width=0.85\linewidth]{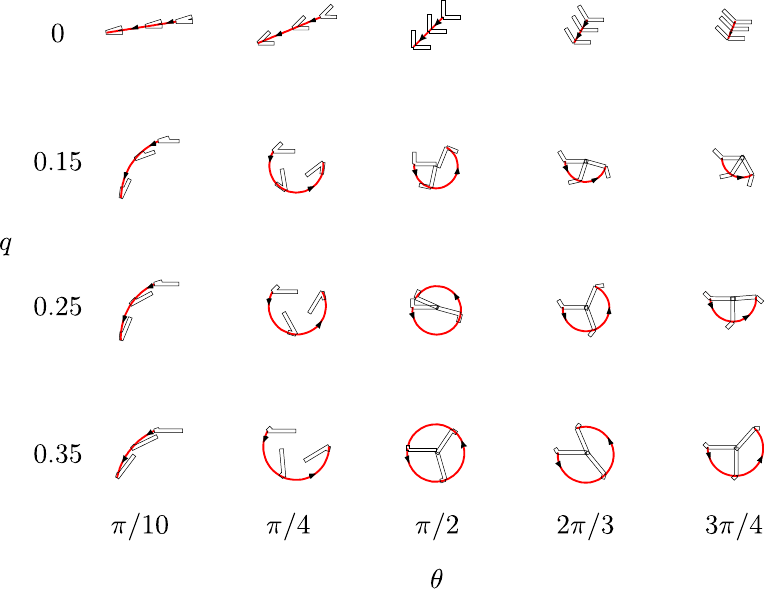}
    \caption{\textbf{Particle trajectories for $j(s) = 1$ and for different $q$ and $\theta$}.  All the trajectories are calculated up to the time period of rotation for the particle with $q=0.35$ and $\theta=\frac{\pi}{2}$.   The trajectories of individual particles are not drawn to a fixed scale; only the total length of each particle is $1$. $\epsilon=10^{-3}$ for all trajectories.}
    \label{fig:3}
\end{figure}

 \subsection{Uniform Surface Flux}  To gain deeper insight into the impact of $q$ and $\theta$ on the motion of particles, we consider the scenario of $j(s)=1$. Under this assumption, the particle motion is solely driven by geometrical asymmetry since the entire particle is uniformly active. In Fig. \ref{fig:2}(c), when the angle between the arms is fixed to $\theta=\frac{\pi}{2}$, $c_s \left(s+\frac{1}{2}-q\right)$ is relatively insensitive to $q$. This is because the concentration at a particular location is dictated by the contributions of the point sources ($\alpha(s)$ defined in Appendix A) spread along the particle. For a fixed angle, the position of the hinge does not have a significant impact on these contributions and thus the concentration is relatively insensitive to $q$. In contrast, for equal arm lengths ($q=0$), a change in $\theta$ has significant impact on $c_s(s)$, see Fig. \ref{fig:2}(d). This occurs because the relative distance between any point on the particle surface with respect to the other arm changes significantly with $\theta$. For smaller angles, the relative distance between the arms decreases and thus the surface concentration is higher. 
 \par{} By employing Eqs. (\ref{eq:unifrm_con_sol}), (\ref{eq:mobility_relation}), and (\ref{eq:steady-state trajectory}), we obtain particle trajectories for a combination of different $q$ and $\theta$ values; see Fig. \ref{fig:3} (also see Supplementary Video 1). We find that when the two arms are of equal lengths, i.e., $q=0$, the particle moves in a straight line irrespective of $\theta$. For $q=0$, due to symmetry, $\Delta c_{0}^{q + \frac{1}{2}} + \Delta c_{0}^{q-\frac{1}{2}}=0$; see Fig. \ref{fig:2}(d). Therefore, the effective torque on the particle, as mathematically described in Eq. (\ref{eq:mobility_relation}), is zero. In addition, the rotational-translational coupling terms also cancel out such that the net turning induced by geometric asymmetry is zero and $\Omega_3=0$. Therefore, the particles only translate. For $q \neq 0$, we observe that the particle turns; see Fig. \ref{fig:3}. As is evident from the figure, the degree and speed of turning depend on $q$ and $\theta$.

\begin{figure}[t]
    \centering
    \includegraphics[width=\textwidth]{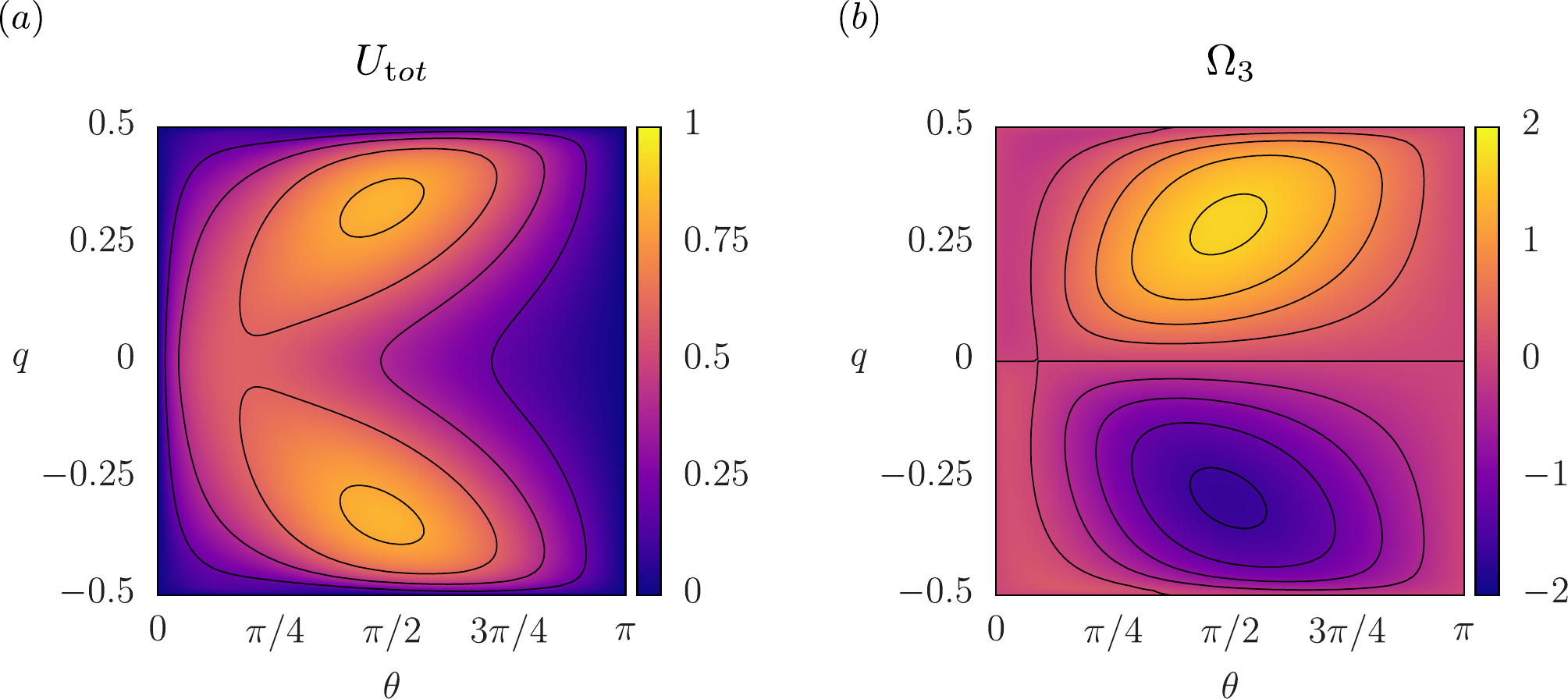}
    \caption{\textbf{Phase plots for $j(s)=1$ and $\epsilon=10^{-3}$}. (a) The speed of the particle $\sqrt{U_1^2+U_2^2}$. (b) The angular velocity of the particle $\Omega_3$.  A maximum in $U_{\textrm{tot}}$ and $\Omega_3$ is observed for $q \approx 0.3$ and $\theta \approx \frac{\pi}{2}$. $\epsilon=10^{-3}$ for both plots.}
   \label{fig:uniform_flx_motion}
\end{figure}

To understand the quantitative dependence of particle trajectory on geometrical parameters, we construct phase plots for $U_{\textrm{tot}}=\sqrt{U_1^2+U_2^2}$ and $\Omega_3$; see Fig. \ref{fig:uniform_flx_motion}. First, we observe that $U_{\textrm{tot}}$ is symmetric about $q=0$, i.e., $U_{\textrm{tot}}(q,\theta)=U_{\textrm{tot}}(-q,\theta)$. This result is  consistent with the expectation, i.e., for $j(s)=1$, switching the positive and negative arms should not influence the speed. In contrast, $\Omega_3(q,\theta)=-\Omega_3(-q, \theta)$, which indicates that the direction of turning switches, but the magnitude remains the same. This dependence agrees with expectation, since changing the sign of $q$ changes the initial orientation, which causes the particle to turn in the opposite direction.      

One of the key features that emerges from Fig. \ref{fig:uniform_flx_motion} is the presence of a maximum in $U_{\textrm{tot}}$ and $\Omega_3$ around $\theta \approx \frac{\pi}{2}$ and $q \approx 0.3$. Therefore, the maximum $U_{\textrm{tot}}$ and $\Omega_3$ is observed for an L-shaped particle with one arm 4 times as long as the other one. We write Eq. (\ref{eq:mobility_relation}) as follows
\begin{equation}
\begin{bmatrix}
    \mathbf{U} \\ \mathbf{\Omega}
  \end{bmatrix} = \mathscr{A} \mathbf{M}(q, \theta) 
  \cdot \begin{bmatrix}
    \mathbf{F}_{\textrm{eff}}(q, \theta) \\ \mathbf{T}_{\textrm{eff}}(q, \theta)
  \end{bmatrix} ,
  \label{eq:hydro_chem}
\end{equation}
where $\mathbf{M}$ is the hydrodynamic mobility and $\mathbf{F}_{\textrm{eff}}$ and $\mathbf{T}_{\textrm{eff}}$ are the effective driving propulsion force and torque.  The geometric parameters $q$ and $\theta$ influence $\mathbf{M}$, $\mathbf{F}_{\textrm{eff}}$, and $\mathbf{T}_{\textrm{eff}}$, but the impact of $q$ and $\theta$ is different on $\mathbf{F}_{\textrm{eff}}$ and $\mathbf{T}_{\textrm{eff}}$. For instance, $\mathbf{F}_{\textrm{eff}}$ is dependent on $\Delta c_0^{q+\frac{1}{2}}$ and $\Delta c_{q-\frac{1}{2}}^{0}$, which are less sensitive to $q$ and more sensitive to $\theta$; see Fig \ref{fig:2}(c,d). $\mathbf{T}_{\textrm{eff}}$ has a similar dependence on $\theta$ but also varies with $q$. In contrast, the mobility parameter $\mathbf{M}$ is more sensitive to $q$ than $\theta$; see Appendix B. The interplay of mobility and driving force yields the optimum values of $q$ and $\theta$.  We note that these optimal values will change if a different $j(s)$ is used, since that will directly impact the values of $\mathbf{F}_{\textrm{eff}}$, and $\mathbf{T}_{\textrm{eff}}$.
\par{} We note that the dependence of $\epsilon$ in Eq. (\ref{eq:hydro_chem}) only comes from $\mathscr{A}$. This occurs because we ignore the higher order corrections in $\mathbf{M}$, $\mathbf{F}_{\textrm{eff}}$ and $\mathbf{T}_{\textrm{eff}}$. To this end, we acknowledge that our analysis will be less applicable when the interactions between the two rods are significant. Additionally, our analysis does not focus on circumferential variation of chemical activity, which could be important in certain scenarios, as discussed in refs. \cite{katsamba2020slender,katsamba2022chemically,rao2018self}. Even with these limitations, our analysis provides a convenient starting point to predict trajectories and design self-propelling composite slender bodies.

\section{Concluding Remarks} \label{sec:conclusion}
\par{} In summary, this article presents a theoretical analysis to predict the two-dimensional motion of self-propelling bent rods. By employing slender body theory and the Lorentz reciprocal theorem, we derive Eqs. (\ref{eq:unifrm_con_sol}), (\ref{eq:mobility_relation}), and (\ref{eq:steady-state trajectory}) to predict the particle trajectories for given $q$, $\theta$, and $j(s)$. Our analysis reveals that the trajectory of the particle is circular, such that the radius is dependent on $q$ and $\theta$, and is insensitive to the slenderness parameter $\epsilon$. The speed and the frequency of the particle are dependent on the values of $q$, $\theta$, and $\epsilon$, and display an optimum with the phase plot of $q$ and $\theta$. The results outlined here provide a convenient method to design self-propellers for micro-bot applications in diagnostics and drug delivery \cite{nelson2010microrobots,bunea2020recent}. For instance, if the propeller is required to travel in a straight line, the propeller should have symmetric arms. If the propeller is required to move and turn faster, optimum values of $q$ and $\theta$ and a smaller $\epsilon$ should be utilized. Physically, our analysis also helps to clarify the role of the particle geometry in hydrodynamic mobility as well as the self-propelling driving force. As such, we can move away from the external force and torque approach, and show that the self-propulsion is driven by the concentration differences across each arm. In the future, our work can be extended to more complex geometries, such as slender bodies with multiple arms and activity profiles with circumferential variation.

\section*{Acknowledgements}
 The authors acknowledge Robert Davis, Bhargav Rallabandi, James Roggeveen, and Howard Stone for their helpful input in the preparation of this manuscript. We would also like to thank Filipe Henrique, Nathan Jarvey, Ritu Raj, and Gesse Roure for their helpful discussions and feedback leading to the culmination of this work. A. Ganguly and A. Gupta also acknowledge the donors of the American Chemical Society Petroleum Research Fund for partial support of this research.

 \section*{Appendix A: Derivation of surface concentration} \label{sec:appendix_b}
 Ignoring any circumferential dependence, we rewrite Eqs. (2) and (3a) in cylindrical coordinates as

\makeatletter 
\renewcommand{\theequation}{A\@arabic\c@equation}
\makeatother
\setcounter{equation}{0}

\begin{subequations}
\begin{equation}
    \begin{aligned}
       \frac{1}{r}\frac{\partial}{\partial r}\left(r\frac{\partial c}{\partial r}\right) + \frac{\partial^2 c}{\partial s^2}=0,\ r>\epsilon,
    \end{aligned} \label{eq:con_non_dim_gov_eq} 
\end{equation}

\begin{equation}
    \epsilon \frac{\partial c}{\partial r}=-j(s),\ r=\epsilon, \label{eq:con_non_dim_bc} 
\end{equation}
\end{subequations}

\noindent and $c \left(r \to \infty \right) \to 0$. To solve our system of equations and obtain the surface concentration $c_s \left(s\right)$, we divide the fluid volume into an inner region and an outer region in the radial direction $\mathbf{e}_r$. In the inner region, we introduce a stretched radial coordinate $\rho = r/\epsilon$. The concentration in the inner region is denoted by $c^i$. Eqs. \eqref{eq:con_non_dim_gov_eq} and \eqref{eq:con_non_dim_bc} in the inner region can be written as,

\begin{subequations}
\begin{equation}
    \begin{aligned}
     \frac{1}{\rho}\frac{\partial}{\partial \rho}\left(\rho \frac{\partial c^i}{\partial \rho}\right)+\epsilon^2 \frac{\partial^2 c^i}{\partial s^2}=0, \ \rho>1,
    \end{aligned}\label{eq:inner_reg_gov_eq}
\end{equation}

\begin{equation}
    \begin{aligned}
      \frac{\partial c^i}{\partial \rho}=-j(s),\ \rho=1.
    \end{aligned}\label{eq:inner_region_boundary_condition} 
\end{equation}
\end{subequations}

\noindent We expand the solute concentration in the inner region in orders of $\epsilon$; $c^i= \log \left(\frac{1}{\epsilon} \right)c^i_0+ c^i_1 + O(\epsilon)$ and solve for the leading order and first order solutions. The resultant concentration profile in the inner region is found to be

\begin{equation}
    c^i=-j(s) \log \rho + K(z;\epsilon) + O(\epsilon^2), \label{eq:inner_sol} 
\end{equation}

\noindent where $K$ in Eq. \eqref{eq:inner_sol} is obtained by matching with the outer solution.

In the outer region, we represent the concentration $c^o$ as being due to a distribution of sources $\alpha(s)$ along an infinitesimally thin space curve along the particle centerline. The position of any point in the fluid volume $\mathbf{x}(r,s)=s\mathbf{e}_t(s) + r \mathbf{e}_r(s)$. The outer region concentration is represented as

\begin{equation}
    c^o=\displaystyle \int_{q-\frac{1}{2}}^{q+\frac{1}{2}} \frac{\alpha(s')}{\left|\mathbf{x}(r,s)-s'\mathbf{e}_t(s')\right|} \ \textrm{d} s'. \label{eq:out_reg_gov_eq} 
\end{equation}

\noindent Note that the concentration field in the outer region, proposed in equation \eqref{eq:out_reg_gov_eq}, decays to zero in the far-field. By using Eq. (1), we evaluate the concentration profile to be

\begin{equation}
    \begin{aligned}
       c^o = \begin{cases}
        \displaystyle \int_{q-\frac{1}{2}}^0 \frac{\alpha(s')}{\left[\left(s-s'\right)^2+r^2\right]^{1/2}} \ \ \textrm{d} s' +  \int_0^{ q +\frac{1}{2}} \frac{\alpha(s')}{\left[\left(s+s' \cos \theta\right)^2+\left(r+s' \sin \theta\right)^2 \right]^{1/2}} \ \ \textrm{d} s' & s < 0, \\
       \displaystyle \int^0_{q-\frac{1}{2}} \frac{\alpha(s')}{\left[\left(s+s' \cos \theta\right)^2+\left(r+s' \sin \theta\right)^2 \right]^{1/2}} \ \textrm{d} s' + \displaystyle \int_0^{q+\frac{1}{2}} \frac{\alpha(s')}{\left[\left(s-s'\right)^2+r^2\right]^{1/2}}  \ \textrm{d} s' & s \geq 0.
       \end{cases}
    \end{aligned}\label{eq:outer_reg_con_sol} 
\end{equation}

\noindent Due to the singularity in Eq. \eqref{eq:outer_reg_con_sol}, we propose a substitution of variables $\zeta=s'-s$ and integrate. For $s \geq 0$

\begin{equation}
    \begin{aligned}
       c^o(r,s\geq0)&=\alpha(s) \displaystyle \int_{-s}^{q+\frac{1}{2}-s} \frac{ \textrm{d} \zeta}{\left[\zeta^2+r^2\right]^{1/2}} + \displaystyle \int_{-s}^{q+\frac{1}{2}-s} \frac{\alpha(s+\zeta)-\alpha(s)}{\left[\zeta^2+r^2\right]^{1/2}} \ \textrm{d} \zeta + \\
       &+\displaystyle \int_{q-\frac{1}{2}-s}^{-s} \frac{\alpha (s+\zeta)}{\left[\left(s+\zeta\right)^2+2\left(s \cos \theta + r \sin \theta\right)\left(s+\zeta\right)+s^2+r^2\right]^{1/2}} \ \textrm{d} \zeta.
    \end{aligned} \label{eq:out_reg_con_leading_order} 
\end{equation}

\noindent Similarly, we can separate out the singularity for $s < 0$. The integrals demonstrate that $c^o(r,s) = 2\alpha(s) \log r + c^o_1 + O(\epsilon)$. To match the outer solution to the inner solution from Eq. \eqref{eq:inner_sol}, we write
$    \lim_{\rho \to \infty} c^i = \lim_{r \to 0} c^o $, which shows $\alpha(s) = \frac{j(s)}{2}$. The surface concentration thus at the slip plane is written as Eq. (4). To evaluate the integrals in Eq. (4), one needs to remove the singularity, as shown in Eq. (\ref{eq:out_reg_con_leading_order}).

 We note that the concentration profiles provided in Fig. \ref{fig:2} display a discontinuity in gradients at the hinge. This is expected since the first-order slender body theory ignores the interaction between the two arms and superposes the corresponding contributions to $c_s$. The interactions are more pronounced when $\theta \rightarrow 0$, and thus our analysis would need to be appropriately corrected.
 
  \section*{Appendix B: Determining the mobility coefficients for a slender bent rod} \label{sec:appendix_c}
  
  The resistance coefficients described Eq. (13) are given as (see Roggeveen and Stone \cite{roggeveen2022motion})
\begin{equation}
 	\begin{bmatrix}
 		A_{11} \\
 		A_{12} \\
 		A_{22} \\
 		\Tilde{B}_{13} \\
 		\Tilde{B}_{23} \\
 		C_{33}
 	\end{bmatrix}=\begin{bmatrix}
 		\frac{1}{8}\left(5-2q+(2q-1) \cos 2\theta\right) \\
 		\frac{1}{8}\left(2q-1\right) \sin 2\theta \\
 		1+\frac{1}{4}\left(2q-1\right) \sin^2 \theta \\
 		-\frac{1}{8}\left(1-2q\right)^2 \sin \theta \\
 		\frac{1}{8}\left(\left(2q+1\right)^2+\left(1-2q\right)^2 \cos \theta \right) \\
 		\frac{1}{12}+q^2
 	\end{bmatrix}.\label{eq:resistance coefficients} 
 \end{equation}
 
 We invert $\mathbf{R}$ and obtain the following mobility coefficients

\begin{equation}
	\begin{bmatrix}
		a_{11} \\
		a_{12} \\
		a_{22} \\
		\Tilde{b}_{13} \\
		\Tilde{b}_{23} \\
		c_{33}
	\end{bmatrix}=\begin{bmatrix}
		\dfrac{-\frac{1}{64} \left(\left(1+2q\right)^2+\left(1-2q\right)^2 \cos \theta\right)^2 + \left(\frac{1}{12}+q^2\right)\left(1+\frac{1}{4}\left(-1+2q\right) \sin^2 \theta\right)}{\frac{1}{768} \left(19+88 q^2-144 q^4 - 12 \left(1-4q^2\right)^2 \cos \theta + \left(1+8 q^2 - 48 q^4\right) \cos 2\theta\right)} \\[25pt]
		\dfrac{-\frac{1}{64}\left(1-2q\right)^2\left(\left(1+2q\right)^2+\left(1-2q\right)^2 \cos \theta\right)\sin \theta - \frac{1}{8}\left(-1+2q\right)\left(\frac{1}{12}+q^2\right)\sin 2\theta}{\frac{1}{768} \left(19+88 q^2-144 q^4 - 12 \left(1-4q^2\right)^2 \cos \theta + \left(1+8 q^2 - 48 q^4\right) \cos 2\theta\right)} \\[25pt]
		\dfrac{\frac{1}{8}\left(\frac{1}{12}+q^2\right)\left(5-2q+\left(-1+2q\right)\cos 2\theta\right) -\frac{1}{64}\left(1-2q\right)^4 \sin^2 \theta}{\frac{1}{768} \left(19+88 q^2-144 q^4 - 12 \left(1-4q^2\right)^2 \cos \theta + \left(1+8 q^2 - 48 q^4\right) \cos 2\theta\right)} \\[25pt]
		\dfrac{\frac{1}{32}\left(-1+2q\right)\left(-3+4q+4q^2+\left(1+2q\right)^2 \cos\theta\right)\sin \theta}{\frac{1}{768} \left(19+88 q^2-144 q^4 - 12 \left(1-4q^2\right)^2 \cos \theta + \left(1+8 q^2 - 48 q^4\right) \cos 2\theta\right)} \\[25pt]
		\dfrac{\frac{1}{64}\left(-\left(\left(\left(1+2q\right)^2+\left(1-2q\right)^2 \cos \theta\right)\left(5-2q+\left(-1+2q\right)\cos 2\theta\right)\right)-2\left(-1+2q\right)^3 \cos \theta \sin^2 \theta\right)}{\frac{1}{768} \left(19+88 q^2-144 q^4 - 12 \left(1-4q^2\right)^2 \cos \theta + \left(1+8 q^2 - 48 q^4\right) \cos 2\theta\right)} \\[25pt]
		\dfrac{\frac{1}{32}\left(17-4q^2+\left(-1+4q^2\right)\cos 2\theta\right)}{\frac{1}{768} \left(19+88 q^2-144 q^4 - 12 \left(1-4q^2\right)^2 \cos \theta + \left(1+8 q^2 - 48 q^4\right) \cos 2\theta\right)}
	\end{bmatrix}\label{eq:mobility_coefficients} 
\end{equation}

\bibliographystyle{ieeetr}
\bibliography{citations}

\end{document}